\documentclass[preprint,preprintnumbers,amsmath,amssymb]{revtex4}
\usepackage{graphicx}
\usepackage{epstopdf}
\usepackage{dcolumn}
\usepackage{bm}
\usepackage{color}

\begin{document}

\title{Decoupling of graphene from Ni(111) via oxygen intercalation}

\author{Yuriy Dedkov,$^{1,2}$ Wolfgang Klesse,$^2$ Andreas Becker,$^2$ Florian Sp\"ath,$^3$ Christian Papp,$^3$ and Elena Voloshina$^{4}$}

\affiliation{$^1$Fachbereich Physik, Universit\"at Konstanz, 78457 Konstanz, Germany}
\affiliation{$^2$IHP GmbH, Im Technologiepark 25, 15236 Frankfurt/Oder, Germany}
\affiliation{$^3$Lehrstuhl f\"ur Physikalische Chemie II, Universit\"at Erlangen-N\"urnberg, Egerlandstr. 3, 91058 Erlangen, Germany}
\affiliation{$^4$Institut f\"ur Chemie und Biochemie - Physikalische und Theoretische Chemie, Freie Universit\"at Berlin, Takustrasse 3, 14195 Berlin, Germany}


\date{\today}

\begin{abstract}
The combination of the surface science techniques (STM, XPS, ARPES) and density-functional theory calculations was used to study the decoupling of graphene from Ni(111) by oxygen intercalation. The formation of the antiferromagnetic (AFM) NiO layer at the interface between graphene and ferromagnetic (FM) Ni is found, where graphene protects the underlying AFM/FM sandwich system. It is found that graphene is fully decoupled in this system and strongly $p$-doped via charge transfer with a position of the Dirac point of $(0.69\pm0.02)$\,eV above the Fermi level. Our theoretical analysis confirms all experimental findings, addressing also the interface properties between graphene and AFM NiO.
\end{abstract}

\maketitle

\section{Introduction}

Graphene (gr), a single layer of carbon atoms arranged in a honeycomb lattice, and its derivatives (flakes, quantum dots, nanoribbons, etc.) attract a lot of attention in the last decade due to its unique electrical and mechanical properties~\cite{CastroNeto:2009,DasSarma:2011br,Meunier:2016fr,Si:2016dc}. However, despite the tremendous success in the lab-based research, the lack of the ``killer'' graphene-based applications limits the spread of the graphene technology in our daily life~\cite{Peplow:2015dd,Park:2016ju}. For example, implementation of graphene in the present-day scalable semiconducting technology requires its synthesis on insulating or semiconducting substrates. Presently this is only possible on h-BN~\cite{Wang:2013do,Tang:2013hy,Tang:2015gn,Driver:2016dx} and Ge~\cite{Wang:2013fq,Lippert:2014fc,Lee:2014dv,Kiraly:2015kaa,Dabrowski:2016im} substrates at the conditions which cannot be easily adapted. Therefore, a new approach to synthesize the graphene/insulator systems was recently developed. It consists of the preparation of graphene on metallic or semiconducting layers via standard chemical vapour deposition (CVD) technique or intercalation under graphene, followed by the oxidation of the underlying metal or semiconductor at relatively high pressure of oxygen. This procedure was used in several recent works, where nearly free-standing graphene on bulk or surface oxide layers was prepared~\cite{Larciprete:2012aaa,Granas:2012cf,Lizzit:2012hh,Omiciuolo:2014dn,Larciprete:2015ek,Voloshina:2016jd}.

This approach can also be used for the fabrication of graphene-protected systems~\cite{Dedkov:2008d,Dedkov:2008e,Sutter:2010bx}. Here as a first example one can consider the synthesis of the exchange biased epitaxial AFM/FM system under graphene, with AFM NiO and FM Ni having Neel and Curie transition temperatures well above the room temperature (AFM: antiferromagnet; FM: ferromagnet). The first attempt to prepare the graphene/NiO/Ni system was undertaken in Ref.~\citenum{Ligato:2016gk} demonstrating the partial oxidation of Ni under graphene with formation of NiO islands at the interface. Neither structural nor electronic properties studies for the gr/NiO/Ni system can be found in the literature up to now.

In the present paper we demonstrate successful complete intercalation of oxygen under graphene on Ni(111). We found that AFM NiO layer is formed at the interface between graphene and the FM Ni. The process of intercalation was monitored with scanning tunneling microscopy (STM) and x-ray photoelectron spectroscopy (XPS). Angle-resolved photoelectron spectroscopy (ARPES) experiments performed on graphene/NiO/Ni(111) showed that graphene is fully decoupled from the underlying substrate and strong $p$-doping of graphene is found with a position of the Dirac point ($E_D$) of $(0.69\pm0.02)$\,eV above the Fermi level ($E_F$). All experimental results are analysed in the framework of the density-functional theory (DFT) approach and good agreement between experimental and theoretical data is found.

\section{Experimental and computational details}\label{Exp_Details}

\begin{figure*}
\includegraphics[width=\linewidth]{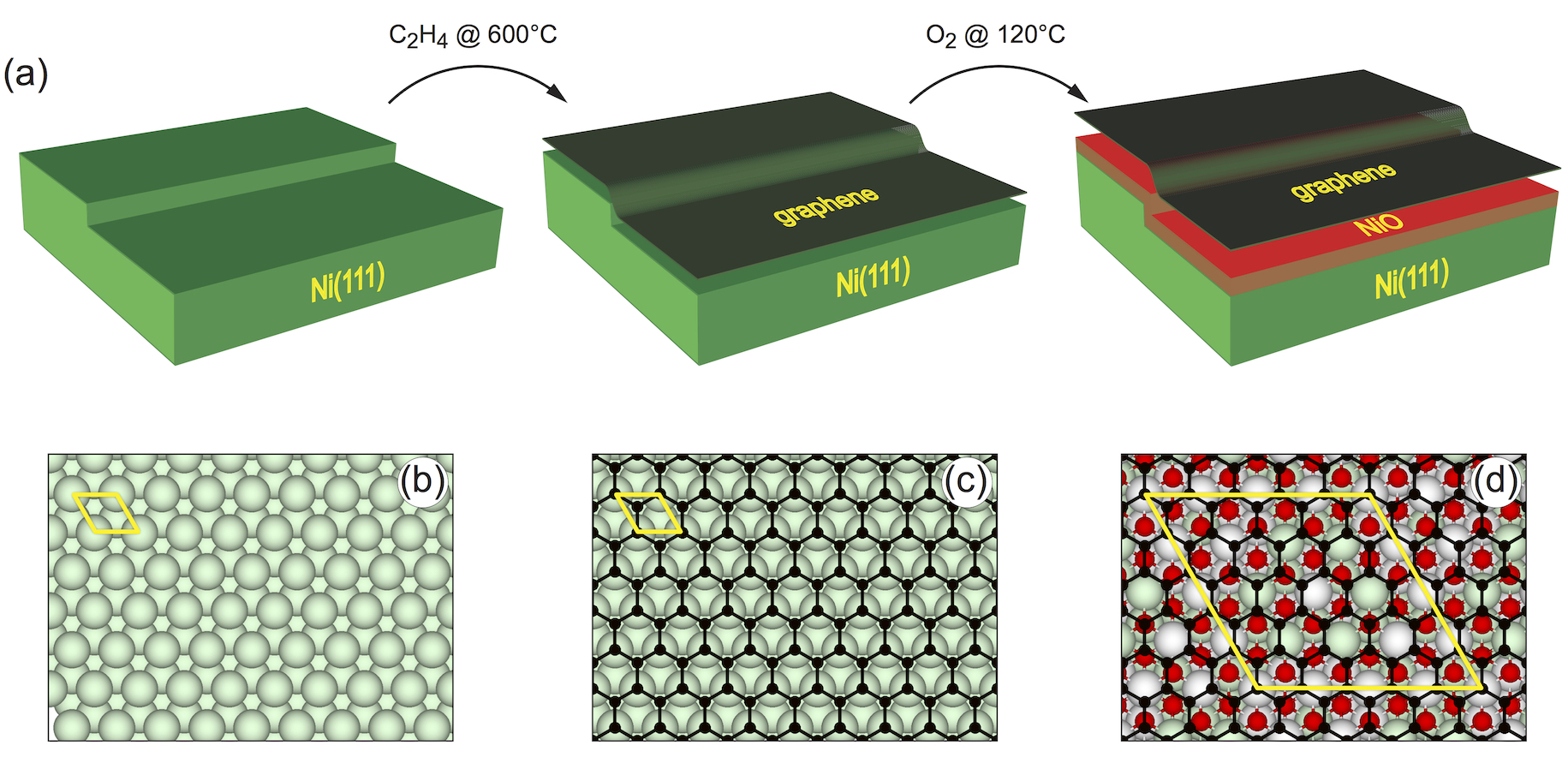}
\caption{(a) Steps of the sample preparation: Ni(111), gr/Ni(111), and gr/NiO$_x$/Ni(111). (b-d) Respective crystallographic structures of the systems from the upper row. Large, middle-size, and small spheres are Ni, O, and C atoms respectively. Yellow rhombuses mark the unit cells of the respective system used in the DFT calculations. In (d) different colours of spheres for Ni correspond to two opposite magnetic moments in AFM NiO.}
\label{scheme}
\end{figure*}

Experiments were performed in two different ultrahigh vacuum (UHV) systems (STM and XPS/ARPES) applying similar sample preparation conditions (Fig.~\ref{scheme}). In both cases, the same Ni(111) single crystal was used as a substrate for graphene growth. Prior to every experiment, Ni(111) was cleaned via cycles of ion-sputtering (Ar$^+$, $800$\,eV, $p=1\times10^{-5}$\,mbar, $15$\,min) and annealing ($750^\circ$\,C, $15$\,min). The cleanness of the surface was verified by means of LEED, STM, or XPS. Graphene was prepared in a usual way via low-pressure CVD procedure using ethene (C$_2$H$_4$) as a precursor ($T=600^\circ$\,C, $p=1\times10^{-6}$\,mbar, $15$\,min). This procedure gives a graphene layer of high-quality with a small fraction of the carbidic phase. The quality of graphene was verified by means of LEED, STM, and XPS/ARPES. Intercalation of oxygen in gr/Ni(111) was performed at $120^\circ$\,C and oxygen was introduced through a stainless steel pipe which end was placed approximately $1$\,mm from the sample surface, that allows to increase drastically the local pressure of oxygen at the sample position (approximately by 2 orders of magnitude). The partial pressure of oxygen in the preparation chamber measured by the hot-cathode ion gauge was $1\times10^{-4}$\,mbar or $5\times10^{-4}$\,mbar. The duration of this procedure determined the amount of intercalated oxygen; however, comparison with the preparations conditions which were used for oxygen intercalation in gr/Ru(0001)~\cite{Voloshina:2016jd} shows that for the formation of the fully decoupled layer of graphene on Ni(111) the oxygen dose has to be at least by factor of 10 higher compared to the former case. This difference is addressed in Sec.~\ref{Results_Discussions}.

All STM experiments were carried out in an UHV system (base pressure $5\times10^{-11}$\,mbar) equipped with an Omicron variable-temperature scanning tunneling microscope. STM measurements were performed in the constant current mode at room temperature using electrochemically etched polycrystalline tungsten tips cleaned in UHV. The sign of the bias voltage corresponds to the voltage applied to the tip. Tunneling current and voltage are labeled $I_T$ and $U_T$, respectively.

The XPS and ARPES experiments were performed at the UE\,56/2 PGM\,1 beamline at the BESSY\,II storage ring (HZB Berlin) in the photoemission station using PHOIBOS 100 2D-CCD hemispherical analyzer from SPECS GmbH. Photon energies are specified for every data set in the text and in the figure captions. In case of ARPES experiments a 5-axis motorized manipulator was used, allowing for a precise alignment of the sample in $k$-space. The sample was pre-aligned (via polar and azimuth angles rotations) in such a way that the tilt scan was performed along the direction perpendicular to the $\Gamma-\mathrm{K}$ direction of the graphene-derived BZ with the photoemission intensity on the channelplate images acquired along the $\Gamma-\mathrm{K}$ direction. The final 3D data set of the photoemission intensity as a function of kinetic energy and two emission angles, $I(E_{kin}, angle1, angle2)$, were then carefully analyzed.

Spin-polarised DFT calculations based on plane-wave basis sets of $500$\,eV cutoff energy were performed with the Vienna \textit{ab initio} simulation package (VASP)~\cite{Kresse:1994,Kresse:1996a,Kresse:1999}.The Perdew-Burke-Ernzerhof (PBE) exchange-correlation functional~\cite{Perdew:1996} was employed. The electron-ion interaction was described within the projector augmented wave (PAW) method~\cite{Blochl:1994} with Ni ($3d$, $4s$), O ($2s$, $2p$), and C ($2s$, $2p$) states treated as valence states. The Brillouin-zone integration was performed on $\Gamma$-centred symmetry reduced Monkhorst-Pack meshes using a Gaussian smearing with $\sigma = 0.05$\,eV, except for the calculation of total energies. For those calculations, the tetrahedron method with Bl\"ochl corrections~\cite{Blochl:1994vg} was used. A $3\times 3\times 1$ k-mesh was used in the case of ionic relaxation and $6\times 6\times 1$ for single point calculation, respectively. The DFT$+U$ scheme~\cite{Anisimov:1997gm,Dudarev:1998vn} was adopted for the treatment of Ni $3d$ orbitals, with the parameter $U_{eff} = U - J = 8$\,eV, which yields a band gap of $4.2$\,eV for the bulk NiO, in good agreement with the experimental value, $4.3$\,eV~\cite{Sawatzky:1984jt}. Dispersion interactions were considered adding a $1/r^6$ atom-atom term as parameterised by Grimme (``D2'' parameterisation)~\cite{Grimme:2006}.  

For the calculation of the electronic properties of the gr/NiO(111) interface the O-terminated NiO(111) surface was modelled by a symmetric slab consisting of thirteen  ionic layers. Graphene layers were adsorbed from both sides of the slab and a vacuum gap is approximately $20$\,\AA. The used supercell has a ($4\times 4$) lateral periodicity with respect to NiO(111) and a ($5\times 5$) lateral periodicity with respect to graphene. During the structural optimisation procedure the seven middle inner layers were fixed at their bulk positions whereas the positions of all other ions were fully relaxed until forces became smaller than $0.02\,\textrm{eV\,\AA}^{-1}$. The band structures calculated for the studied systems were unfolded to the graphene ($1\times 1$) primitive unit cell according to the procedure described in Ref.~\citenum{Medeiros:2014ka} with the \texttt{BandUP} code. All calculations for gr/Ni(111) were performed according to procedure from Refs.~\citenum{Voloshina:2014jl,Voloshina:2011NJP,Voloshina:2013cw}.

\section{Results and discussions}\label{Results_Discussions}

\begin{figure*}
\includegraphics[width=0.7\linewidth]{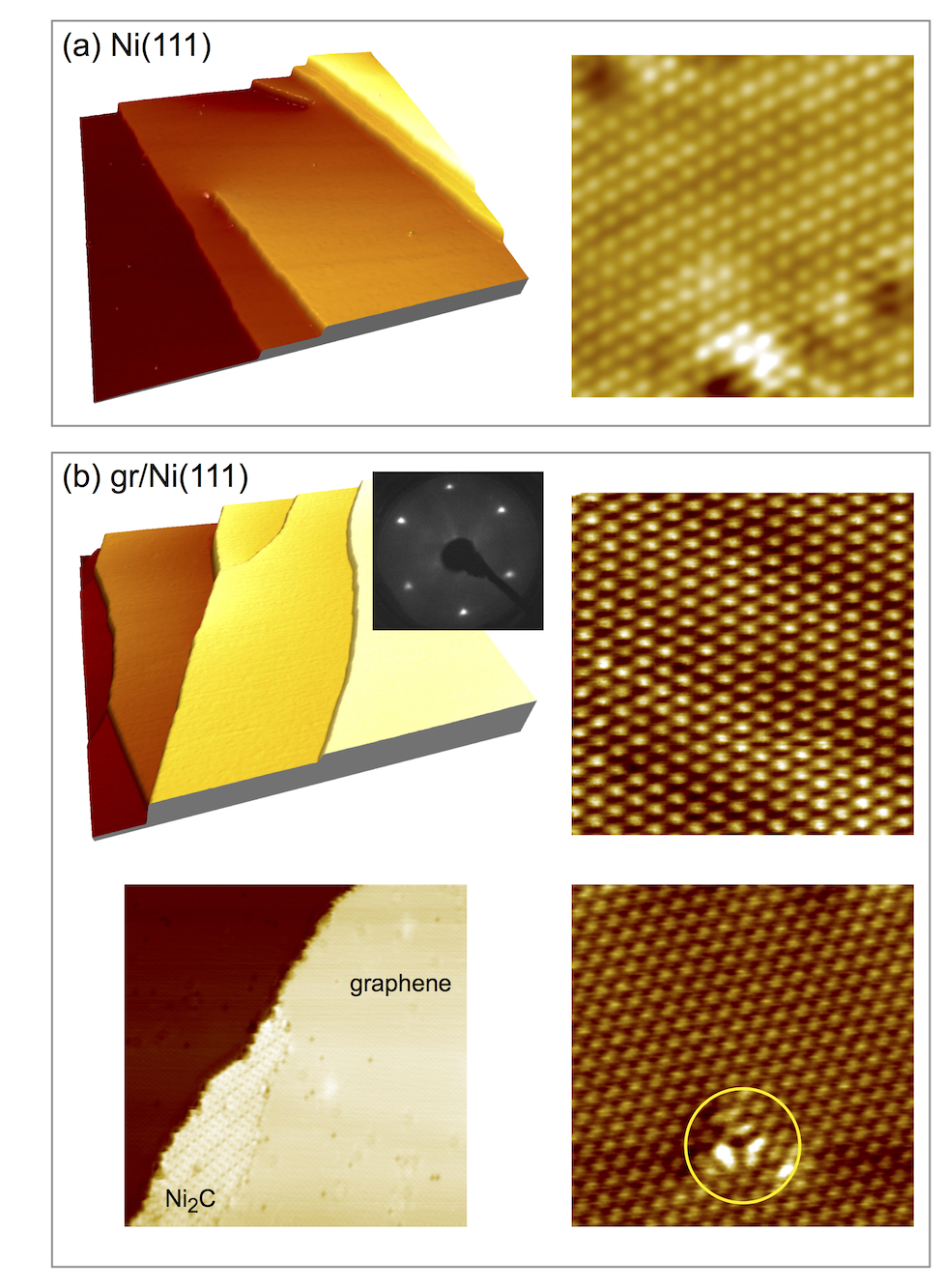}
\caption{(a) Large scale and atomically-resolved STM images of Ni(111). Scanning parameters: (left) $175 \times 175\,\mathrm{nm}^2$, $U_T=+400$\,mV, $I_T=1$\,nA; (right) $3.5 \times 3.5\,\mathrm{nm}^2$, $U_T=+50$\,mV, $I_T=5$\,nA. (b) Large scale and atomically-resolved STM images of gr/Ni(111). Lower row shows representative defects found on gr/Ni(111): carbidic phase (left) and missed-atom single defect (right). Scanning parameters for the upper row: (left) $210 \times 210\,\mathrm{nm}^2$, $U_T=-100$\,mV, $I_T=5$\,nA; (right) $4 \times 4\,\mathrm{nm}^2$, $U_T=-100$\,mV, $I_T=10$\,nA. Scanning parameters for the lower row: (left) $28 \times 28\,\mathrm{nm}^2$, $U_T=+100$\,mV, $I_T=10$\,nA; (right) $4.5 \times 4.5\,\mathrm{nm}^2$, $U_T=+20$\,mV, $I_T=10$\,nA. Inset in (b) shows a LEED image of gr/Ni(111) taken at $120$\,eV of the primary electron energy.}
\label{STM_Ni111_grNi111}
\end{figure*}

\textit{STM}. Intercalation of oxygen in gr/Ni(111) and formation of the gr/NiO/Ni(111) system was verified using real-space STM as well as by spectroscopic methods. Figure~\ref{STM_Ni111_grNi111}(a) shows STM images of the clean Ni(111) surface, demonstrating ordered surface on the large (left) and small (right) scale. After preparation of graphene on Ni(111) [Fig.~\ref{STM_Ni111_grNi111}(b)], the large scale STM images show well ordered graphene layer with small fraction of the carbidic phase (Ni$_2$C)~\cite{Lahiri:2011iu,Jacobson:2012be} and very small number of single-atom vacancy defects~\cite{Wang:2013jx}. Two STM images showing such imperfections of the gr/Ni(111) system are presented in the bottom row of panel (b). These defected areas of the graphene layer (missing carbon atoms or/and borders between graphene and carbidic phases) can be considered as areas where intercalation of oxygen can take places. The high-quality of gr/Ni(111) is also confirmed by LEED images of this system, one of which is shown as an inset in Fig.~\ref{STM_Ni111_grNi111}(b). It is interesting to note that in our real-space STM experiments we observe the decrease of the number of screw-dislocations on the Ni(111) surface after preparation of a graphene layer, that can be assigned to the reduction of the surface (or interface) energy in the system and the respective atom rearrangement upon graphene formation. 

\begin{figure*}
\includegraphics[width=0.8\linewidth]{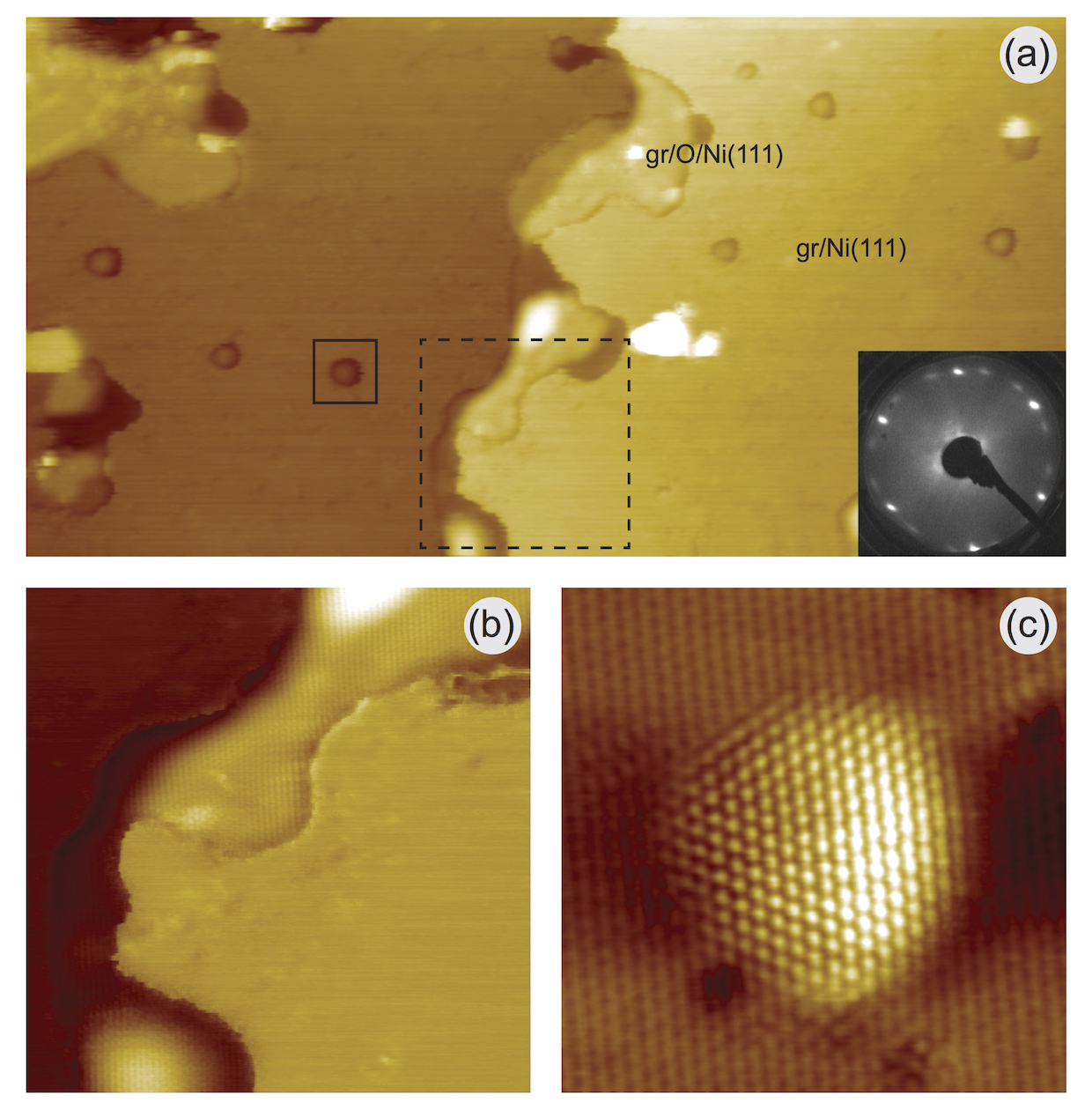}
\caption{(a) Large scale STM image of partially intercalated oxygen in gr/Ni(111). Scanning parameters: $116 \times 60\,\mathrm{nm}^2$, $U_T=-100$\,mV, $I_T=5$\,nA.  Dashed- and solid-line squares in (a) mark areas which zoomed STM images are shown in panels (b) and (c), respectively. Scanning parameters: (b) $20 \times 20\,\mathrm{nm}^2$, $U_T=-100$\,mV, $I_T=5$\,nA; (c) $6 \times 6\,\mathrm{nm}^2$, $U_T=-20$\,mV, $I_T=5$\,nA. Inset in (a) shows a LEED image ($100$\,eV) obtained after fully intercalated oxygen in gr/Ni(111).}
\label{STM_grONi111}
\end{figure*}

Intercalation of oxygen in gr/Ni(111) was performed at high partial pressure of gas and $T=120^\circ$\,C. In order to trace the intercalation process and directly compare systems before and after intercalation, we present results for not fully decoupled graphene layer on Ni(111) (Fig.~\ref{STM_grONi111}). Panel (a) shows a large scale STM image acquired after keeping gr/Ni(111) at $120^\circ$\,C and $p(O_2)=1\times10^{-4}$\,mbar for $30$\,min. After this procedure we can clearly discriminate between sample areas where oxygen was intercalated or not. It is interesting to note, that intercalation takes places mainly around step edges of the sample, where graphene might be partially lifted. Also several places on the terraces where oxygen intercalates in gr/Ni(111) and forms areas of the lifted graphene can be identified. The zoomed STM images of the step edge and the area on the terrace where intercalation takes place are shown in panels (b) and (c), respectively. Here we can conclude that intercalation of oxygen (and probably other species in the previous experiments reviewed in the literature) in graphene on metals takes places around extended or point defects in the graphene layer. For example one of such point defects is visible in the lower left part of image in panel (c). On other places, where graphene forms a perfect layer on the metallic substrate, it behaves as a protection carpet keeping the underlying metal surface intact. 

\textit{XPS and TPD}. In order to perform decoupling of the full graphene layer on Ni(111), much higher oxygen dose was used: $120^\circ$\,C, $p(O_2)=5\times10^{-4}$\,mbar for $60$\,min. Comparing the oxygen intercalation in gr/Ni(111) (present work) and in gr/Ru(0001)~\cite{Voloshina:2016jd} we can conclude that much higher oxygen dose is necessary in the former case. Such difference can be assigned to the different crystallographic structures of a graphene layer on metallic substrates. In case of gr/Ni(111), graphene forms a flat layer on a lattice matched Ni(111) surface with the distance to the top Ni layer of $2.089$\,\AA. For graphene on Ru(0001) a strongly buckled moir\'e structure is formed, due to the lattice mismatch between graphene and Ru lattices, where alternating places of the short ($2.10$\,\AA) and long ($3.37$\,\AA) C-Ru distances can be found. Such buckled structure with places of a locally lifted graphene might be easily intercalated by oxygen or other species, compared to flat gr/Ni(111), that is observed in our experiments.

\begin{figure*}
\includegraphics[width=0.7\linewidth]{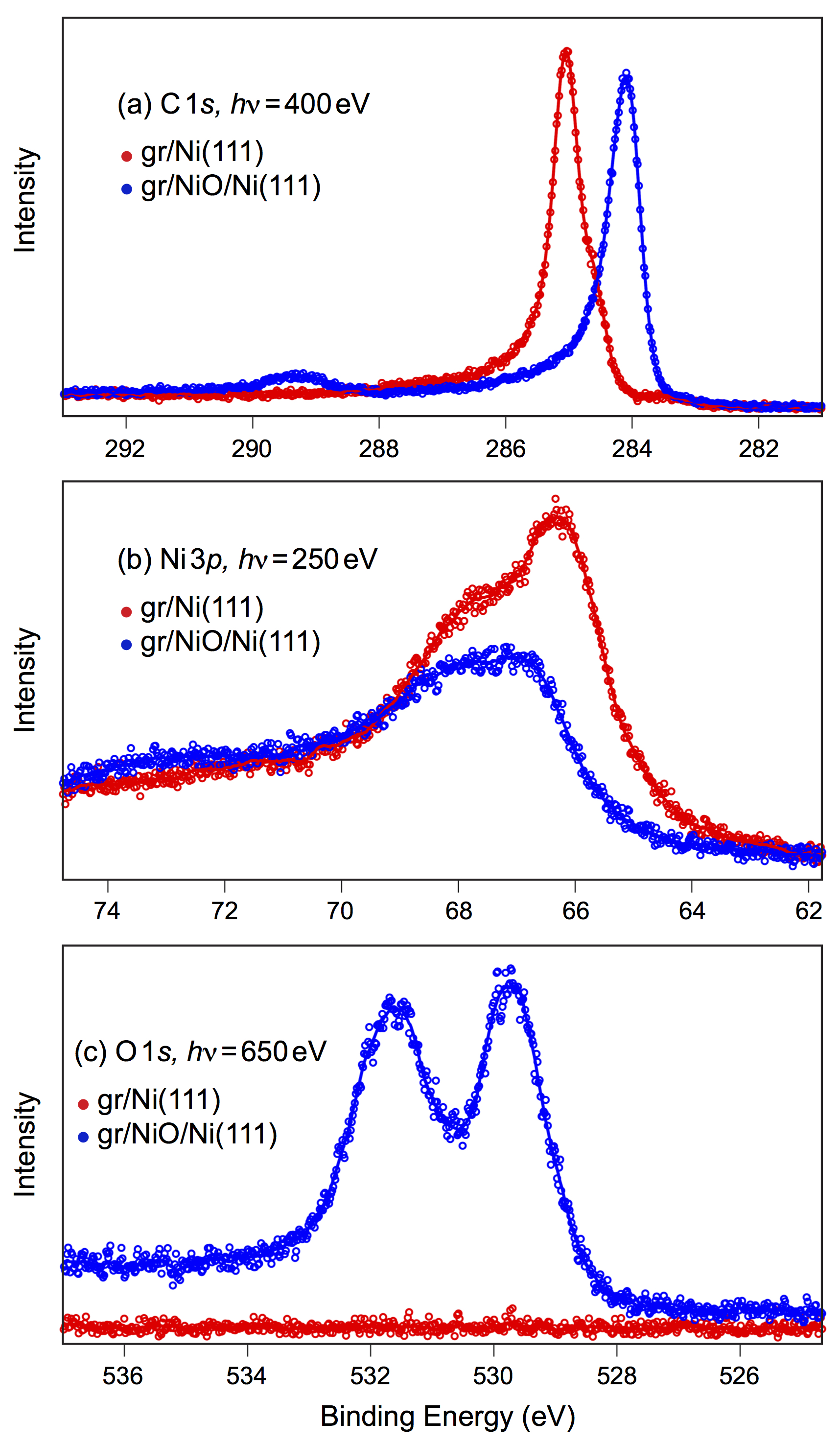}
\caption{XPS spectra acquired before and after full oxygen intercalation in gr/Ni(111): (a) C\,$1s$, $h\nu=400$\,eV, (b) Ni\,$3p$, $h\nu=250$\,eV, (c) O\,$1s$, $h\nu=650$\,eV.}
\label{XPS_BESSY_grNi_grONi111}
\end{figure*}

The chemical state of elements and the electronic structure of gr/Ni(111) and gr/NiO$_x$/Ni(111) were studied by means of XPS and ARPES. Fig.~\ref{XPS_BESSY_grNi_grONi111} shows XPS spectra collected before and after intercalation of oxygen in gr/Ni(111). The C\,$1s$ XPS spectrum of gr/Ni(111) [Fig.~\ref{XPS_BESSY_grNi_grONi111}(a)] is in very good agreement with previously published data~\cite{Dedkov:2008e,Weser:2010,Voloshina:2011NJP,Spath:2016db}: the binding energy of this line is $285.05$\,eV; the shoulder observed at the smaller binding energies can be assigned to the areas of gr/Ni(111) with implanted Ar atoms~\cite{Spath:2016db} (originates from the preparation procedure of the Ni crystal). The Ni\,$3p$ XPS spectrum of gr/Ni(111) [Fig.~\ref{XPS_BESSY_grNi_grONi111}(b)] shows a clear expected unmodified spin-orbit doublet.

The intercalation of the molecular oxygen in gr/Ni(111) at the conditions described earlier leads to the clear modifications of all XPS emission lines. The binding energy of the C\,$1s$ line in the intercalation system is $284.09$\,eV indicating the change of the doping level of graphene compared to the parent gr/Ni(111) system. In this case, we can expect the decoupling of the electronic states of graphene from those of the Ni(111) substrate due to the formation of the NiO layer at the interface. Absence of any residual XPS signal, which can be assigned to the gr/Ni(111) system, indicates the full decoupling of a graphene layer from the substrate and restoring of its nearly free-standing character. Intercalation of oxygen leads to the oxidation of the interface nickel and formation of the thin layer of NiO as can be deduced from the surface sensitive Ni\,$3p$ XPS spectra of gr/NiO/Ni(111) [Fig.~\ref{XPS_BESSY_grNi_grONi111}(b)], which shows a very good agreement with the data obtained for the bulk NiO (energy position as well as the satellite structure at larger binding energies)~\cite{Sunding:2010gr}.

\begin{figure*}
\includegraphics[width=\linewidth]{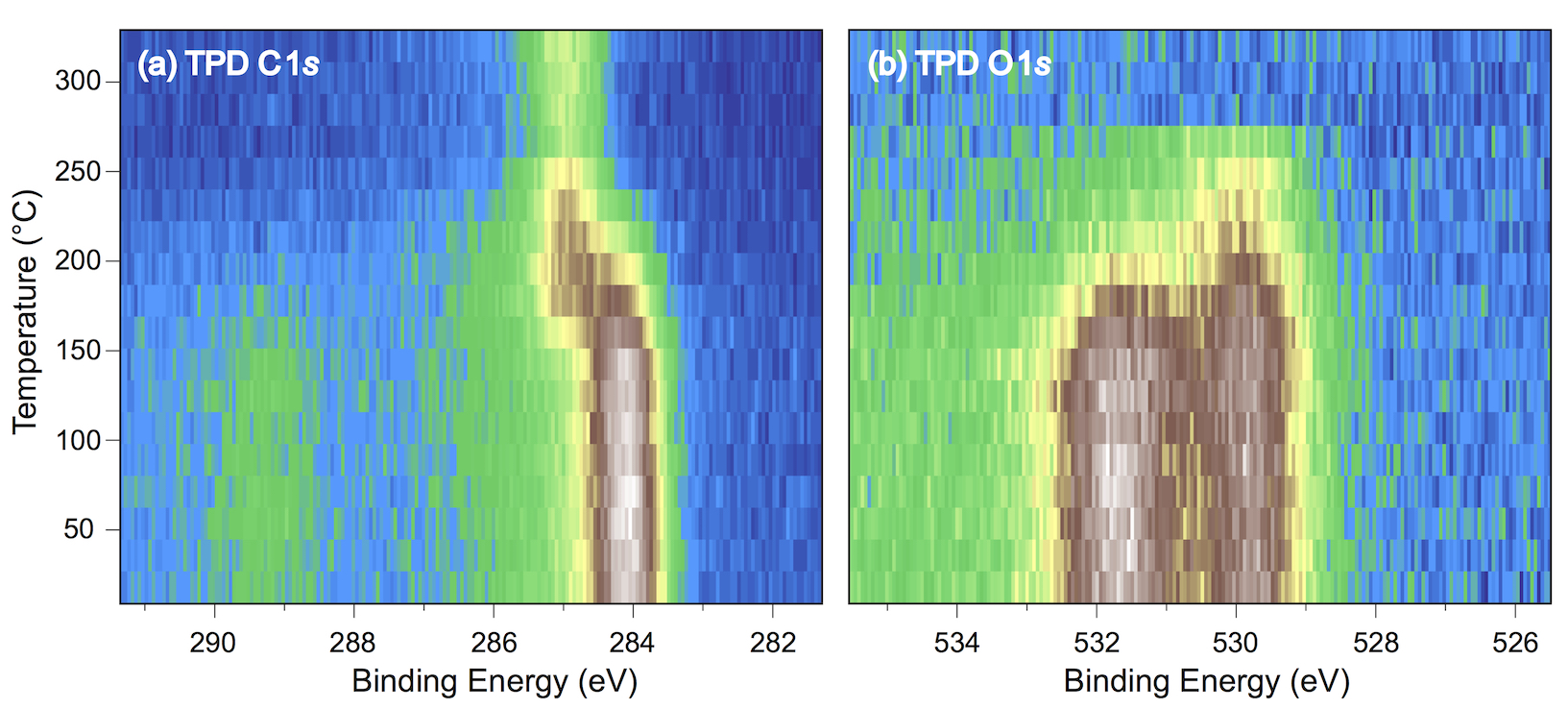}
\caption{Photoemission intensity maps as a function of binding energy and sample temperature for (a) the C\,$1s$ and (b) O\,$1s$ emission lines recorded in the TPD-XPS experiments for the graphene/NiO$_x$/Ni(111) system.}
\label{XPS_TPD_C1s_O1s}
\end{figure*}

XPS measurements of the O\,$1s$ line for the gr/NiO/Ni(111) system reveal two spectral components. The low binding energy component at $529.71$\,eV can be clearly assigned to the emission from O$^{2-}$ in NiO~\cite{Lorenz:2000cx,Rettew:2011kd}. The second one at $531.61$\,eV might be due to the presence of the OH$^-$ groups on the surface~\cite{Rettew:2011kd,Tyuliev:1999je} or can be connected with defects in the oxide layer~\cite{Roberts:1984du}. Our thermo-programmed desorption XPS (TPD-XPS) experiments performed for the C\,$1s$ and O\,$1s$ XPS lines (Fig.~\ref{XPS_TPD_C1s_O1s}) suggest that higher binding energy O\,$1s$ component is due to OH$^-$ groups. It is due to the fact that emission in the C\,$1s$ line assigned to COOH-groups ($289.3$\,eV)~\cite{Stankovich:2006dr} and the discussed  second component in the O\,$1s$ line disappear simultaneously as the sample temperature approaches $150^\circ$\,C. Further increase of the sample temperature leads to the ``burning'' of graphene via reaction of carbon atoms with oxygen from the NiO layer. Some weak C\,$1s$ emission can still be detected, however, no complete layer of graphene can be observed (as concluded from the mesurement of very weak dispersion of the graphene $\pi$ bands in the ARPES experiments).

\begin{figure*}
\includegraphics[width=\linewidth]{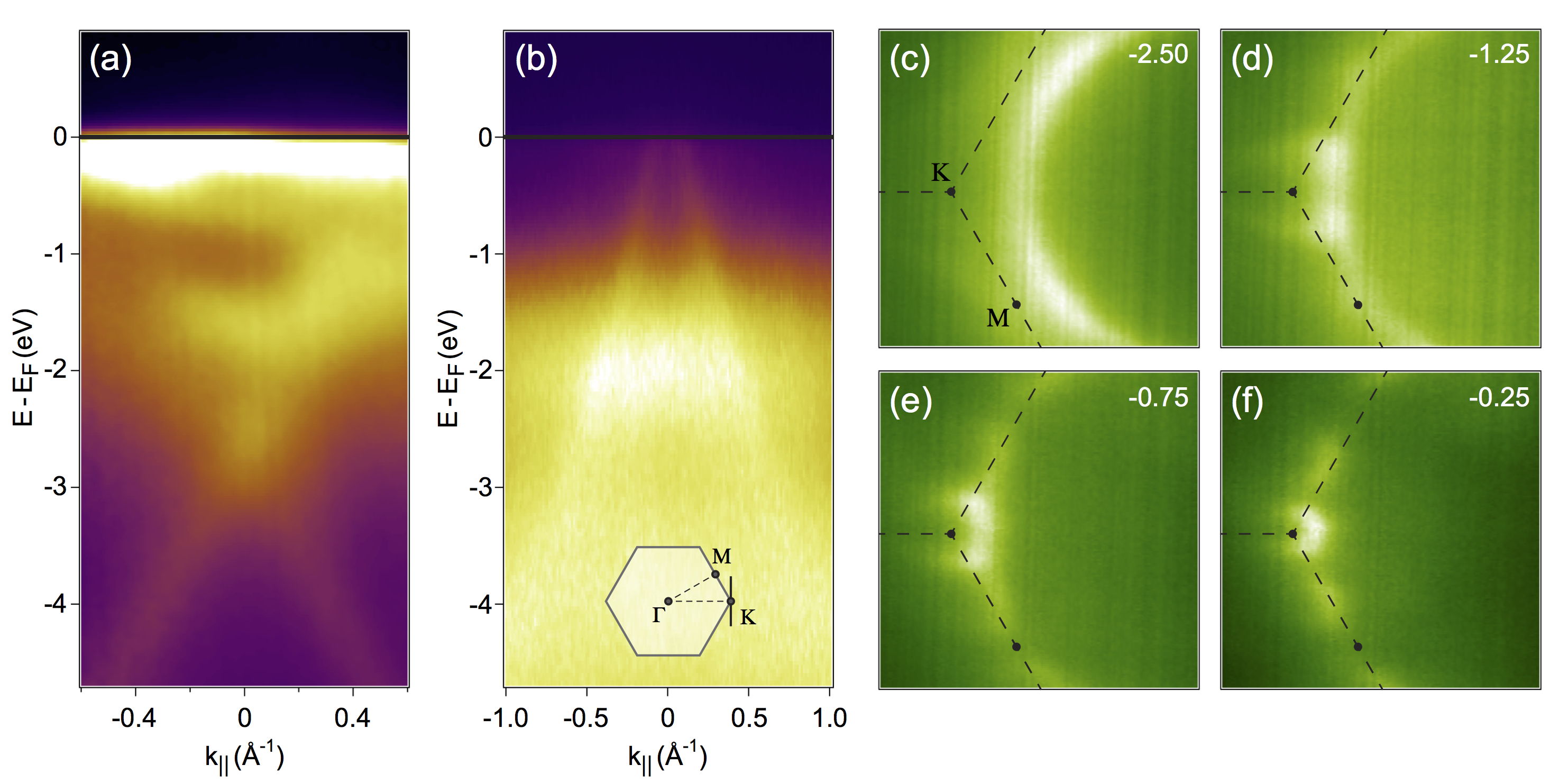}
\caption{ARPES intensity maps for (a) gr/Ni(111) and (b) gr/NiO$_x$/Ni(111) collected along the direction in the BZ of graphene perpendicular to the $\Gamma-\mathrm{K}$ direction (see inset of (b)). (c-f) Constant binding energy cuts extracted at the respective values of $E-E_F$(eV) marked in every panel. Dashed lines and the capital letters mark the BZ of graphene and the respective high-symmetry points. Photon energy in all experiments is $h\nu=65$\,eV.}
\label{ARPES_grNi_grONi}
\end{figure*}

\textit{ARPES}. Figure~\ref{ARPES_grNi_grONi} compiles the results of our ARPES studies of gr/Ni(111) and gr/NiO/Ni(111). All images were extracted from the complete 3D data sets collected through the desired portions of the Brillouine zone of the studied systems. Panels (a) and (b) show the ARPES intensity maps for gr/Ni(111) and gr/NiO/Ni(111) as a function of the binding energy and the wave vector along the direction perpendicular to $\Gamma-\mathrm{K}$ of the graphene Brillouine zone (marked by the solid line in the inset of (b)).

Considering the ARPES map of gr/Ni(111) (Fig.~\ref{ARPES_grNi_grONi}(a)), the strong modification of the electronic structure of a graphene layer is observed. As discussed in a series of previous works~\cite{Voloshina:2011NJP,Bertoni:2004,Dedkov:2010jh,Voloshina:2014jl}, the strong doping of a graphene layer and \textit{hybridization} of the Ni\,$3d$ and graphene $\pi$ valence band states lead to the shift of the graphene $\pi$ bands by $\approx2.5$\,eV below $E_F$ at the $\mathrm{K}$ point, where a series of Ni\,$3d$-C\,$p_z$ \textit{hybrid} states is formed. This effect leads to the strong redistribution of the valence band states in graphene and Ni and, e.\,g., induced magnetism in graphene layer was detected~\cite{Weser:2010}, which was also later confirmed for other gr/FM interfaces~\cite{Matsumoto:2013eu,Weser:2011,Usachov:2015kr,Marchenko:2015ka}.

Intercalation of oxygen in gr/Ni(111) leads to the drastic changes in the observed ARPES picture [Fig.~\ref{ARPES_grNi_grONi}(b)]. The electronic structure of a graphene layer in the vicinity of $E_F$ is fully restored; graphene $\pi$ bands demonstrate linear dispersion; strong $p$-doping of graphene is observed with a position of the Dirac point at $E_D-E_F=0.69$\,eV. Valence band states of the underlying NiO having mainly Ni\,$3d$ character are located at $E-E_F\approx-2$\,eV and $E-E_F\approx-4.45$\,eV. From the absence of the visible \textit{hybridization} between graphene $\pi$ and NiO valence band states we can conclude that one of the conditions for this effect is broken, namely the space overlap of the Ni\,$3d$ and C\,$p_z$ states. Therefore we can conclude that the NiO layer underneath graphene is terminated by oxygen.

Constant energy cuts extracted at different binding energies from the ARPES data set for gr/NiO/Ni(111) are shown in Fig.~\ref{ARPES_grNi_grONi}(c-f) (the respective binding energies are marked in every panel). The presented picture is characteristic for the free-standing graphene demonstrating a circular shape of the Fermi surface and clear trigonal warping of the band at high binding energies. These data demonstrate clear ARPES replicas which are rotated by $15^\circ$ with respect to the main bands and this observation correlates with the LEED spots observed for this system (inset of Fig.~\ref{STM_grONi111}(a)), where additional LEED patterns are visible. This effect can be connected with the appearance of misoriented graphene areas due to the relatively large lattice mismatch between graphene and NiO(111) (see discussion below).

\begin{figure*}
\includegraphics[width=\linewidth]{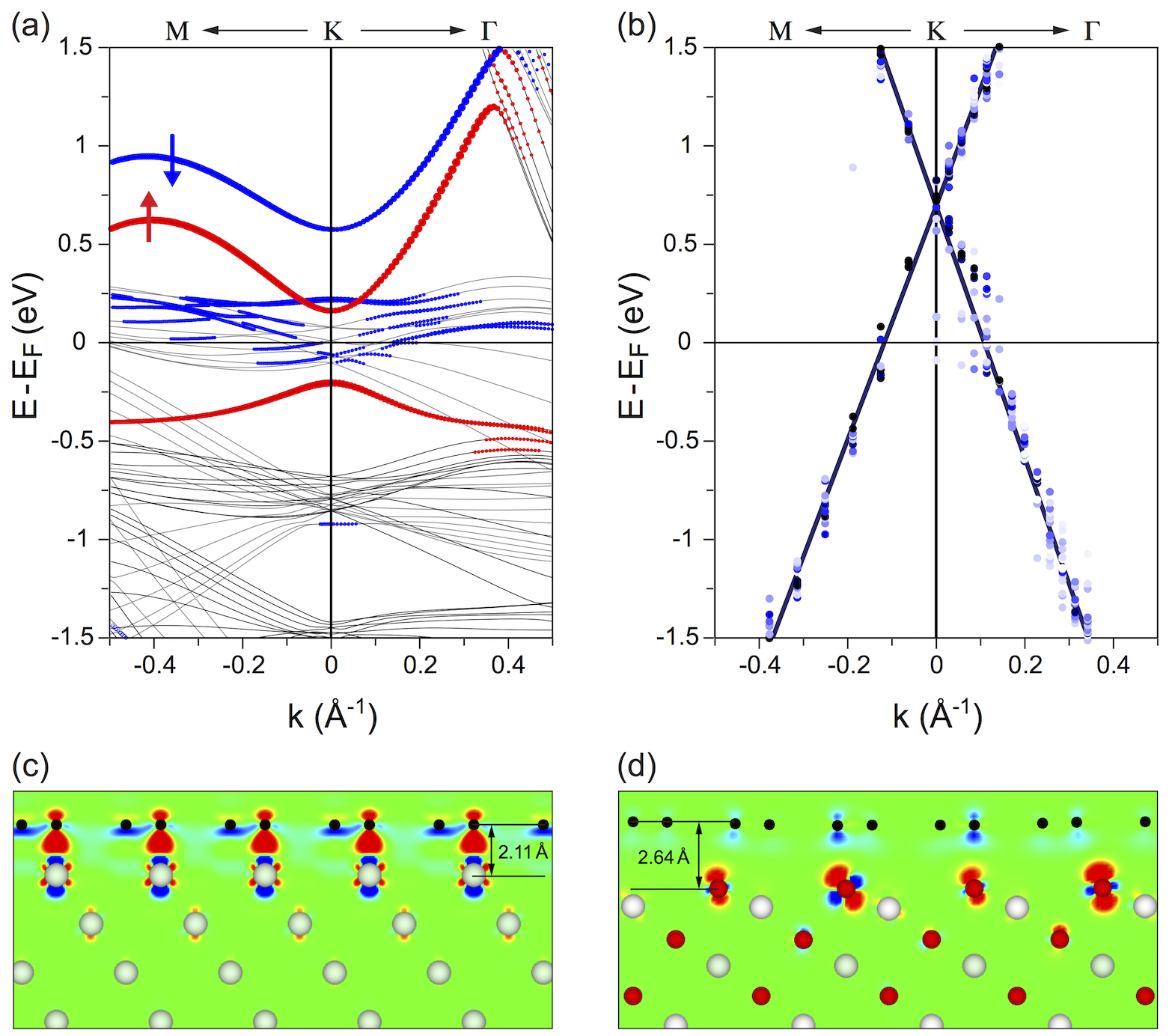}
\caption{Band structures of (a) gr/Ni(111) and (b) gr/NiO(111) in the vicinity of the $\mathrm{K}$ point of the graphene-derived BZ. The weight of the graphene-derived $p_z$ character is highlighted by the size of filled circles superimposed with the plot of the band structure. Grey coloured bands are Ni\,$3d$ states. Spin-up (red) and spin-down (blue) states of graphene are marked in (a). In (b) lines are used for the estimation of the graphene Dirac point. (c,d) Calculated difference electron density, $\Delta \rho(r)=\rho_{gr/s}(r)-\rho_{gr}(r)-\rho_{s}(r)$, for gr/Ni(111) and gr/NiO(111), respectively ($s$ - substrate). $\Delta \rho$ is colour coded as red ($+7\,e/\mathrm{nm}^3$) -- green ($0$) -- blue ($-7\,e/\mathrm{nm}^3$).}
\label{DFT_grNi_grONi}
\end{figure*}

\textit{DFT}. The electronic structure of gr/NiO(111)/Ni(111) was studied in the framework of the DFT approach. Structural optimizations show that graphene-substrate distance is increased from $2.11$\,\AA\ for gr/Ni(111) to $2.64$\,\AA\ for gr/NiO(111). As stated in Sec.~\ref{Exp_Details} the band structure of the lattice-matched gr/Ni(111) interface was calculated for the $(1\times1)$ unit cell and the one for gr/NiO(111) was calculated in the supercell geometry. The corresponding results are shown in Fig.~\ref{DFT_grNi_grONi} (a) and (b). The presented band structure for gr/Ni(111) is in perfect agreement with previously published data~\cite{Voloshina:2011NJP,Bertoni:2004,Voloshina:2014jl,Weser:2011}, where one can clearly see that the original band structure of graphene is fully destroyed and a series of the hybrid Ni-C interface states is formed. 

After oxygen intercalation in gr/Ni(111), a thin layer of NiO is formed underneath graphene and this layer is oxygen terminated. Therefore in our DFT calculations graphene was placed on the O-terminated NiO(111) slab in the supercell geometry. The resulting band structure of graphene in this system, which was unfolded on the $(1\times1)$ unit cell of graphene is shown in Fig.~\ref{DFT_grNi_grONi} (b). [The plotted discrete band structure is due to the unfolding procedure and complicated unit cell of the gr/NiO(111) interface.] The restoring of the original linear dispersion of the $\pi$ states in the band structure of the free-standing-like graphene layer is clearly visible. Our DFT calculations show that graphene in the gr/NiO system is strongly $p$-doped with the position of the Dirac point at $E-E_F\approx0.72$\,eV, that is in a very good agreement with the value extracted from the experimental data. The effect of decoupling of graphene from the substrate after oxygen intercalation is also demonstrated in Fig.~\ref{DFT_grNi_grONi} (c) and (d), where electron density difference $\Delta \rho$ is presented for gr/Ni(111) and gr/NiO(111), respectively. In agreement with the band structures one can see that in the former case the strong hybridization between Ni\,$3d$ and graphene\,$\pi$ states is observed. In case of the gr/NiO(111) interface, a graphene layer is fully decoupled from the substrate and only the effect of the charge transfer, i.\,e. the $p$-doping of graphene, is observed. 

\section{Conclusions}

In our work we demonstrate the successful intercalation of molecular oxygen in gr/Ni(111) and formation of a graphene-protected AFM/FM system. This process leads to the formation of a thin NiO layer as deduced from our experimental data. In the gr/NiO interface, a graphene layer is fully decoupled from the substrate, demonstrating nearly free-standing character. The effect of adsorption of graphene on NiO leads to the $p$-doping of graphene with the position of the Dirac point of $0.69$\,eV above the Fermi level. All experimental results are confirmed by state-of-the-art DFT calculations demonstrating very good agreement between all data.

\section*{Acknowledgement}

We thank HZB for the allocation of synchrotron radiation beamtime. F.S. and C.P. thank the German Research Foundation (DFG) for funding through SFB 953 ``Synthetic Carbon Allotropes'' and the Cluster of Excellence ``Engineering of Advanced Materials''. E.V. thanks DFG for financial support within the Priority Programme SPP 1459 ``Graphene'' and the North-German Supercomputing Alliance (HLRN) for providing computer time.


\clearpage

\clearpage

\clearpage

\clearpage

\clearpage

\clearpage

\clearpage

\end{document}